\documentclass[12pt]{JHEP3}
\usepackage{mathrsfs}
\usepackage{amsmath}
\usepackage{epsfig}
\title{ The zeros and poles of the partition function}
\author{Jingbo Wang \\
  Department of Applied Physics,
 Xi'an Jiaotong University,
 Xi'an, 710049, People's Republic of China\\
 \email{ shuijing31@gmail.com }}

\abstract { In this paper, we consider the physical meaning of the
zeros and poles of partition function. We consider three different
systems, including the harmonic oscillator in one dimension, Riemann
zeta function and the quasinormal modes of black hole.  }

\keywords{ Partition function, Riemann zeta function, complex
analysis, quasinormal mode} \preprint{} \dedicated{} \maketitle
\begin{document}
\section{Introduction}
The partition function plays a fundamental role in physics. In
statistics mechanics, if we know the partition function of the
system, we can calculate all thermal quantities of this system. In
quantum field theory, we can get the correlate function from the
generating function. Even in AdS/CFT correspondence, the dictionary
relate the partition functions of two sides. Usually the partition
function is real function of real variables, and is positive.

A good understanding of the thermodynamics properties of a system
can be obtained by studying the complex zeros of its partition
function. For example, the famous Lee-Yang theorem\cite{LY} said
that the zeros of the partition function of ferromagnetic spin-1/2
Ising model with two-spin interaction lie on the imaginary H axis.
This imply that the ferromagnetic Ising model cannot have a phase
transition in a finite (real) magnetic field.

The zeros of the partition function occur in other place, such as in
Riemann zeta function\cite{R1}, and in quasinormal modes of the
black hole\cite{Be}. In this paper, we want to investigate what we
can learn about those systems from the zeros(poles) of their
partition functions.
\section{Three systems}
First let's consider the simplest system, the 1D harmonic
oscillator. The energy level of the system are
\begin{equation}\label{1}
   E_n  = (n + 1/2)\hbar \omega  = (n + 1/2)E_0.\hskip 5mm
   n=0,1,2,\cdots
\end{equation}
The partition function is
\begin{equation}\label{2}
Z(\beta ) = \sum\limits_{n = 0}^\infty  {\exp ( - \beta E_n )}  =
\frac{{\exp ( - 1/2\beta E_0 )}} {{1 - \exp ( - \beta E_0 )}}.
\end{equation}
Obviously the partition function has poles at $\beta_n=2\pi i
n/E_0$, and from complex analysis we know that this partition can be
expressed as follows:
\begin{equation}\label{3}
    Z(\beta ) = \frac{{\exp ( - 1/2\beta E_0 )}} {{\beta E_0
}}\prod\limits_{n \ne 0} {(1 - \frac{{\beta E_0 }} {{2\pi ni}})\exp
( - \frac{{\beta E_0 }} {{2\pi ni}})}  = \frac{{\exp ( - 1/2\beta
E_0 )}} {{\beta E_0 }}\prod\limits_{n > 0} {(1 + \frac{{\beta ^2 E_0
^2 }} {{4\pi ^2 n^2 }})}.
\end{equation}
That is, the energy levels and the poles are dual to each other. In
this simple system, we observe the relation $\Delta E\Delta \beta =
2\pi i.$ But on the other hand, if the energy changed into
$E_n=nE_0$, the poles are unchanged, so the poles alone can't
determine the energy level.

Next let's consider the Riemann zeta function,
\begin{equation}\label{4}
    \zeta (s) = \sum\limits_{n = 1}^\infty  {\frac{1} {{n^s }}}  =
\prod\limits_{p(prime)} {\frac{1} {{1 - p^{ - s} }}}.
\end{equation}
This function can be considered as the partition function of the
Connes-Bost system\cite{CM} or the primon gas system\cite{Ju}. The
energy level of those system is $E_n=ln(n),n=1,2,\cdots$On the other
hand, from the zeros of the zeta function, we can get,
\begin{equation}\label{5}
    \zeta (\beta ) = \sum\limits_{n = 1}^\infty  {\exp ( - \beta \ln n)}
= \exp (\frac{{\gamma  + \ln \pi }} {2}\beta  - \ln 2)\frac{1}
{{\beta  - 1}}\prod\limits_\rho  {(1 - \frac{\beta } {\rho })}
\prod\limits_{n > 0} {(1 + \frac{\beta } {{2n}})} \exp ( -
\frac{\beta } {{2n}}).
\end{equation}
where $\gamma$ is the Eular constant, $\rho$ are the nontrivial
zeros of the zeta function, and the famous Riemann hypothesis state
that the real part of all $\rho$ is 1/2.

Then we get another dual system, the energy level
$E_n=ln(n),n=1,2,\cdots$ and the zeros $\rho_n$. This duality can
also be seen from the Riemann-Weil explicit formula relating the
prime numbers p and the imaginary part of the Riemann zeros
$\rho_n$. This duality is similar to Selberg's duality\cite{Se}
between the lengths of the primitive orbits and the eigenvalues of
the Laplace-Beltrami operator on a compact Riemann surface with
negative curvature. We also have the celebrated Selberg's trace
formula.

Consider a system with the energy level $E_n$ of the form
$\rho_n=1/2+iE_n$. In Berry and Keating's opinion\cite{BK}, this
system correspondence to a quantum chaos system, and ${lnp}$ are
closed periods orbits for this system, ${lnn}$ the pseudoperiods.

We investigate the other system, the quasinormal modes(QNMs) of
black hole physics. The QNMs can be used to get the one-loop
correction for quantum gravity in AdS/CFT correspondence\cite{De}.
But we will show that we can get more then just the one-loop
correction from the QNMs. In a semiclassical quantization of gravity
the partition function can be written schematically as
\begin{equation}\label{6}
    Z = \sum\limits_{g*} {\det ( - \nabla _{g*}^2 )^{ \pm 1} \exp ( -
S_E (g*))}.
\end{equation}
Here $g*$ are saddle points of the Euclidean gravitational action
$S_E$. In this paper, we just consider the case which there is only
one such saddle point. In paper \ref{De} it was derived that the
determinant in \ref{6} can be formula in terms of the quasinormal
mode of the spacetime, or
\begin{equation}\label{7}
    Z_B  = \exp (Pol(\Delta ))\prod\limits_{z*} {\frac{{\sqrt
{z*\overline {z*} } }} {{2\pi T}}} \prod\limits_{n \geqslant 0} {(n
+ \frac{{iz*}} {{2\pi T}})^{ - 1} (n - \frac{{i\overline {z*} }}
{{2\pi T}})^{ - 1} }.
\end{equation}
where $z*$ are QNMs.

 In our opinion, we conjecture that the QNMs are
zeros or poles of the full quantum gravity partition function. Then
the partition function can be written as
\begin{equation}\label{8}
Z = \exp ( - S_E (g*))\prod\limits_{z*} {(1 - \frac{z} {{z*}})}.
\end{equation}
Though the zeros only contribute to the one-loop correction, from
above examples we know that those zeros are related to the energy
level (or periods orbits) of this system. They are dual to each
other. In many case, the QNMs are equally spaced asymptotic. And
this fact may uncover that the energy level of the quantum gravity
are equally spaced asymptotic, just as the area eigenvalues in loop
quantum gravity, \[ A = 8\pi \gamma \hbar G\sum\limits_n {\sqrt {j_n
(j_n  + 1)} }.
\]
\section{Conclusion}
The partition function determine the properties of the system, and
the poles and zeros of the partition function partly determine the
partition function, though not completely. In this paper, we
investigate three system: the 1D harmonic oscillator, the Riemann
zeta function and the quasinormal modes of black hole. For those
system, we know some properties from the mathematical or physical
point of view independently. But if we combine those two view, maybe
we can get more information. \acknowledgments{This work was partly
done at Beijing Normal University. This research was supported in
part by the Project of Knowledge Innovation Program (PKIP) of
Chinese Academy of Sciences, Grant No. KJCX2.YW.W10 }
\bibliography{zeta1}
\end{document}